\documentclass[twocolumn,aps,pra]{revtex4}\usepackage{epsfig}
\usepackage{graphics}\usepackage{amsthm}

\begin{document}
\title{Quantum non-Gaussianity from a large ensemble of single photon emitters}

\author{Luk\' a\v s Lachman and Radim Filip}

\affiliation{Department of Optics, Faculty of Science, Palack\' y University,
17. listopadu 1192/12,  771~46 Olomouc, Czech Republic}

\email{lachman@optics.upol.cz, 
  filip@optics.upol.cz} 


\begin{abstract}
Quantum attributes of light have been related to non-classicality so far, i. e. to incompatibility with mixtures of coherent states. The progress in quantum optics indicates that this feature does not suffice to witness exotic behavior of light. Contrary, quantum non-Gaussianity is starting to appear as a promising and applicable property reflecting  interesting states of light that are suitable for quantum protocols. We investigate a newly introduced hierarchy of criteria of quantum non-Gaussianity and predict this attribute can be observed on light emitted from many single photon emitters, even if the light undergoes realistic optical losses. It contrasts negativity of Wigner function, that is unreachable in experimental platforms with losses above fifty percent.
\end{abstract}

\maketitle

\section{Introduction}
Quantum optics has introduced a new view on light according to which light consists of indivisible quanta of energy called photons \cite{Einstein}.  This concept  becomes more striking, if quantum features are observed on light with high energy, because these states achieve fields considered so far as a domain of classical optics. A generally accepted meaning of quantum behavior in optics defines non-classical states as states beyond mixtures of coherent states \cite{Glauber}. Many experiments indicate that non-classicality can approach a truly macroscopic limit. A typical example is suppressing of shot noise by bright two mode squeezed vacuum \cite{Bondani,Masha,Masha2} or sub-Poissonian light yielded from heralding on these states \cite{Haderka,Leuchs, Silberhorn}. Recently, robust non-classicality of light from many single photon emitters has been predicted \cite{me}. All these steps take us closer to a preparation of building stones of quantum optics - to Fock states \cite{Cooper, Hofheinz}. These states have unique quantum features and non-classicality is only a first step in identifying them. The quantum non-Gaussianity is a stricter tool for recognising them.
It has been introduced recently a methodology of criteria of quantum non-Gaussianity that witnesses  more precisely the discrete character of light. Quantum non-Gaussianity donates incompatibility of a state with any mixture of squeezed coherent states \cite{Genoni1,Genoni,Radim, me2,Huntington}. Because this feature excludes a more general set of states, it differs fundamentally from non-classicality and imposes a more appropriate condition on required quantum features. For example, this test can be used as a witness of sufficiency of quantum channel for single photon quantum key distribution \cite{Mikolaj}.

The quantum non-Gaussianity criteria represent a higher threshold for quality of many photon sources. On the other hand, quantum non-Gaussianity is still very loss-tolerant \cite{me2}, as has been experimentally verified for single photon states \cite{Ivo, Ulrik}.  A pioneering measurement in the diagnostics of multi-photon states attributed the quantum non-Gaussianity to a state of light, which was occupied up to four photons at average. The light was conditionally generated by spontaneous parametric down conversion \cite{nature}. $N$ temporal modes of a single photon state were measured jointly. The challenging was to suppress sufficiently the higher (meaning more than $N$) photon contributions so that the quantum non-Gaussianity was still observable. In this paper, we discuss the possibilities of the newly introduced hierarchy of quantum non-Gaussianity to witness this feature on different experimental platforms, where a stream of photons is radiated from many independent single photon emitters.

\section{The hierarchy of criteria}
A multi-channel detector can be used for a test of quantum non-Gaussianity. It splits an impinging light into spatially or temporarily separated channels that are detected by sensitive single photon detectors \cite{Mogilevtsev,Zambara,walmsley,Hradil}. The mostly available single photon detector is an avalanche photodiode (APD), which distinguishes with a low efficiency only signal from vacuum. If the multi-channel detector consists of $n+1$ channels, it discriminates simultaneous $n$ clicks irrespective to the last one from an error event when signal is registered in all channels.  The knowledge of the detector provides to build a hierarchy of operational criteria that incorporate the detector even with its imperfections as unbalanced splitting or variation of quantum efficiency in individual channels. These criteria test, whether a probability of an error event is suppressed so much that any quantum Gaussian state cannot exhibit such click statistics.

\begin{figure}
\centerline {\includegraphics[width=10cm]{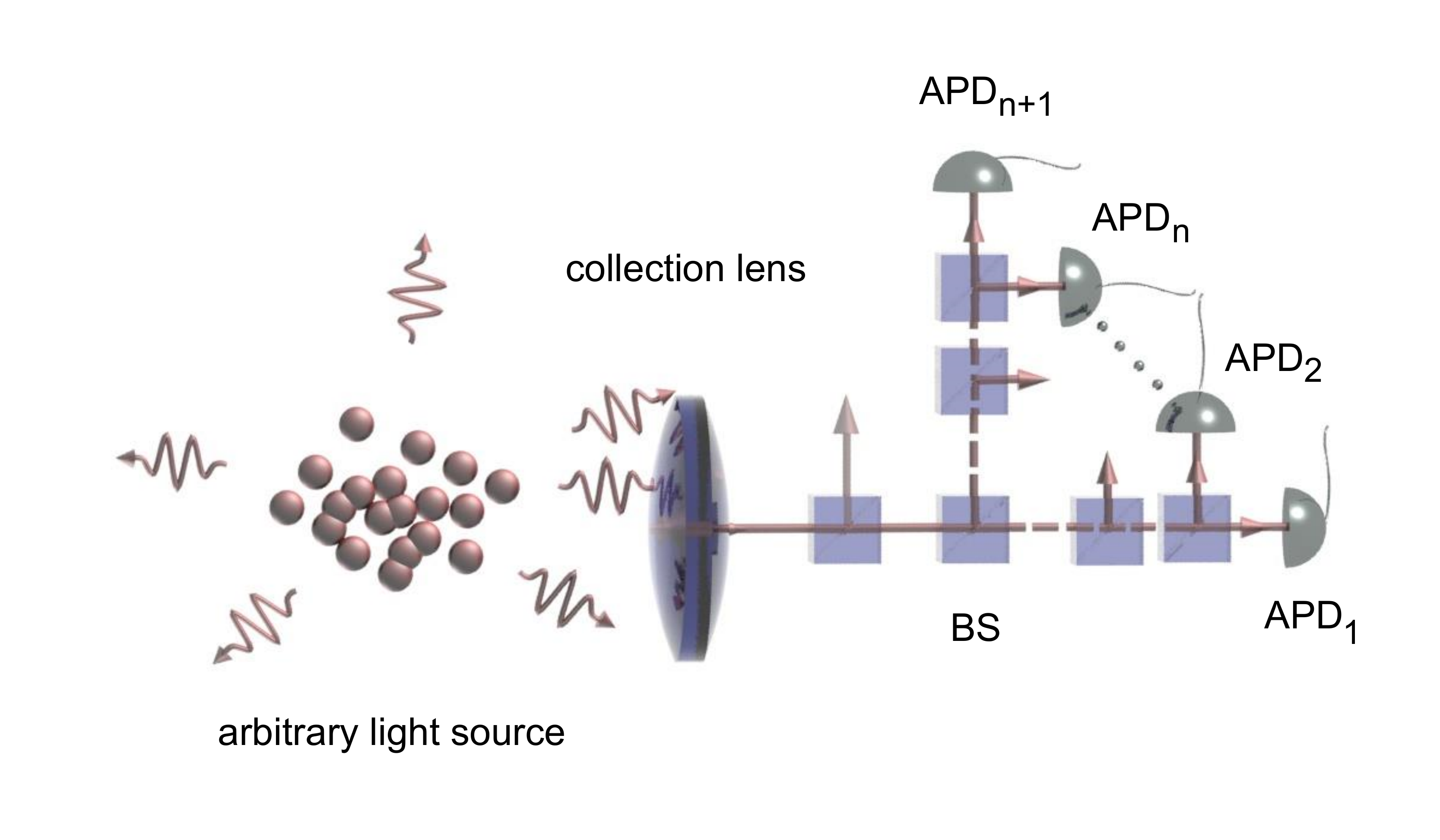}}
\caption{Experimental set-up for testing quantum non-Gaussianity of light. A light source consisted of stable amount of single photon emitters radiates light, that goes through attenuator to a multi-channel detector. The observation of quantum non-Gaussianity is exerted by an array of beam splitters and avalanche photodiods, that provide to test incompatibility of clicks statistics of the source with any mixture of Gaussian states.}
\label{expSet}
\end{figure}

Let us define the relevant probabilities as $R_n$ (probability of success) and $R_{n+1}$ (probability of error). The success means that $n$ expected clicks are registered (irrespective to the event on the last APD), the error tells about presence of unwanted $n+1$ clicks. The source fails when it produces less than $n$ clicks. These probabilities can be combined in a linear form
\begin{equation}
F_{a,n}(\rho)=R_n(\rho)+a R_{n+1}(\rho),
\label{functional}
\end{equation}
where $a$ is a free parameter and $F_{a,n}$ is a functional linear in state $\rho$. The criterion of quantum non-Gaussianity is derived from optimizing this functional over all mixtures of squeezed coherent states. Due to the linearity of the functional, the maximum belongs to a set of pure states - a set of squeezed coherent states.
These states are determined by their minimal variance $V$ and by a complex amplitude $\alpha$ that holds an angle $\phi$ with the direction of minimal squeezing \cite{Yuen}. It means that optimizing the functional in Eq. (\ref{functional}) corresponds to searching for a maximum of a function with three real inputs 
\begin{equation}
F_n(a)=\max F_{a,n}(\vert \alpha \vert, \phi, V),
\end{equation}
where $F_n(a)$ can not be overcome by any mixture of squeezed coherent states for a given $a$. The witnessing of quantum non-Gaussianity corresponds to an appropriate choice of the parameter $a$, that gives 
\begin{equation}
R_n+a R_{n+1}>F_n(a).
\label{nonGACond}
\end{equation}
It suffices to find just a single $a$ satisfying (\ref{nonGACond}) to witness quantum non-Gaussianity. The optimal choice of the parameter leads to the wanted threshold on success probability $R_n$ as a function of $R_{n+1}$ \cite{Radim}.
In general, numerical optimization has to be used to find the threshold. Only if the light is occupied by a very small mean number of photons, the threshold for any number of detectors gains an approximate analytic formula 
\begin{equation}
R_{n}^{n+2}>H_n^{4}(x) \left[ \frac{R_{n+1}}{2 (n+1)^3}\right]^{n},
\label{approx}
\end{equation}
where $H_n(x)$ is the Hermite polynomial of order $n$ and $x$ satisfies $H_{n+1}(x)=0$.
 Figure~2 plots the thresholds obtained numerically for different orders and compare them with approximations in Eq. (\ref{approx}). Details of the formulations and derivations of this task are published in Ref. \cite{nature}.
 
 \begin{figure}
\centerline{\includegraphics[width=9cm]{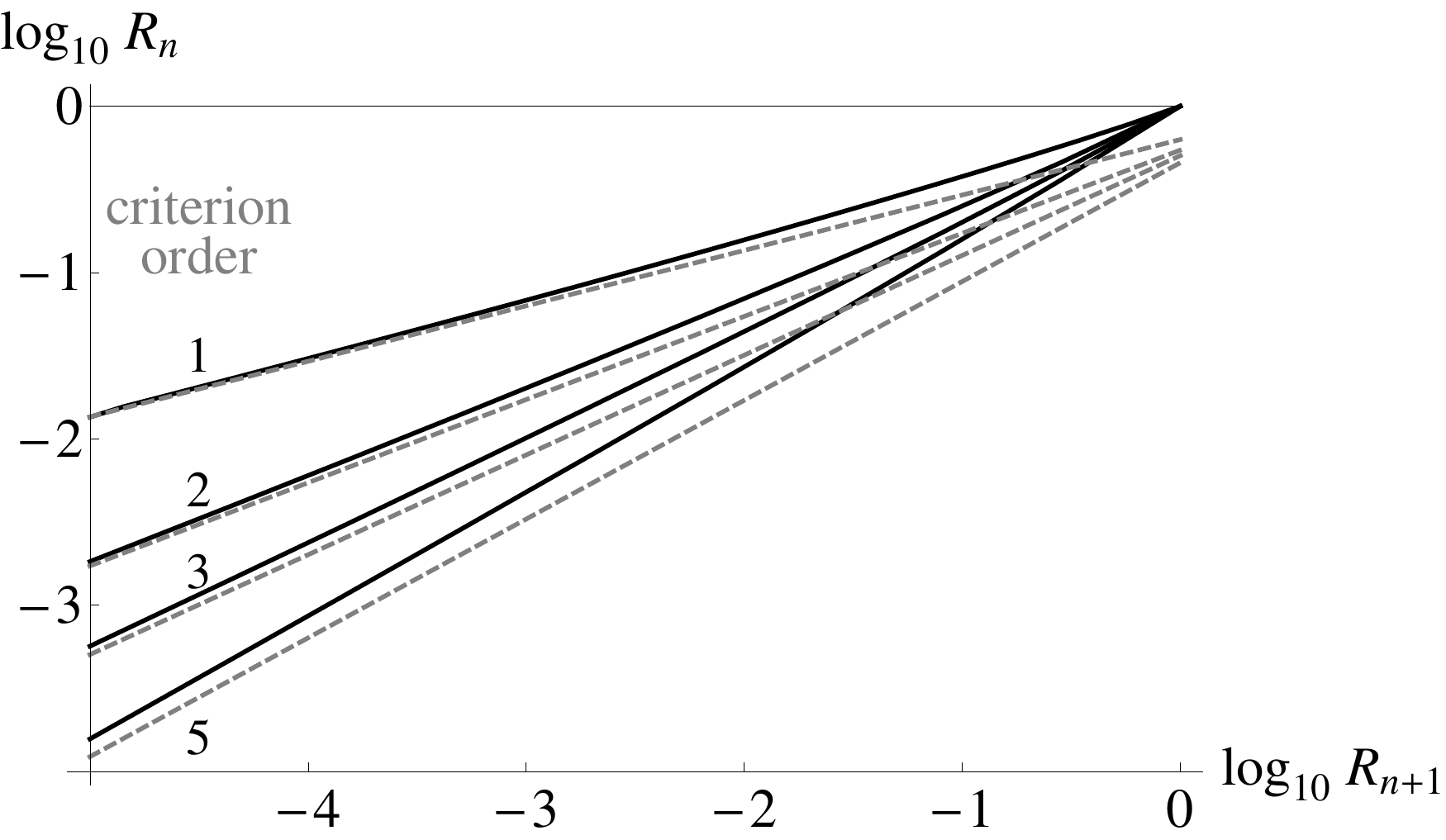}}
\caption{The thresholds of quantum non-Gaussianity corresponding to different criteria in the hierarchy are depicted by solid lines. The dashed lines appropriate to the best linear approximation in the limit of strongly attenuated states.}
\label{gaussTh}
\end{figure}
 
\section{Model of a realistic source}
At many experimental platforms, a test of multi-photon quantum non-Gaussianity can exploit $m$ independent emitters, that radiate just a single photon with typically low probability. Expectedly, the best example is an ensemble of trapped ions acting as two-level systems \cite{me}. In this case, a single two-level system produces no multi-photon light. The density matrix of the emitted light gains
\begin{equation}
\rho=\left[\eta \vert 1 \rangle \langle 1 \vert + (1-\eta)\vert 0 \rangle \langle 0 \vert \right]^{\otimes m},
\label{model}
\end{equation}
where $\vert n \rangle$ is Fock state and $\eta$ is overall efficiency of emission and detection of a photon from a single emitter. The statistics of clicks yielded by this state in a multi-channel detector is determined by a probability of no-clicks $R_{0,n}$, meaning $n$ chosen detectors do not register any signal. We assume here the symmetrical multichannel detector, where light is divided to all channel equally. However, our methodology can be directly used for any known structure of the multi-channel detector. This structure can be obtained by measurement and detector tomography \cite{Hradil,tomo,banaszek,Paul,Franson} using coherent light. 
For the considered model in Eq. (\ref{model}), the no-clicks probabilities get
\begin{equation}
R_{0,k}=\left(1-\frac{\eta k}{N}\right)^m,
\label{R0S}
\end{equation}
where $N$ is a total number of channels in the detector.
The relation between no clicks probabilities and clicks probabilities is given by an identity
\begin{equation}
R_n=1+\sum_{k=1}^n{n \choose k}(-1)^k R_{0,k},
\label{RnS}
\end{equation}
where $R_n$ qualifies a probability a chosen group of $n$ detectors registers $n$ clicks. The expressions in Eqs. (\ref{R0S}) and (\ref{RnS}) enable to analyze a test of quantum non-Gaussianity on this simple model.

Because $m$ single photon emitters can not radiate more than $m$ photons in this ideal case, the criterion of order $m$ witnesses the quantum non-Gaussianity of this state for arbitrary efficiency $\eta$. A criterion with lower order can detect it from certain efficiency $\eta$ as well. Figure~3 shows the dependence of this threshold efficiency on the order of the criterion and number of emitters. We can reduce therefore the complexity of multi-channel detector only if collection efficiency is high. This results should further stimulate experimental development of multi-channel detectors \cite{Hradil,tomo,banaszek,Franson} and their tomographic characterization \cite{Hradil,tomo,banaszek,Franson}. 

\begin{figure}
\centerline{\includegraphics[width=9cm]{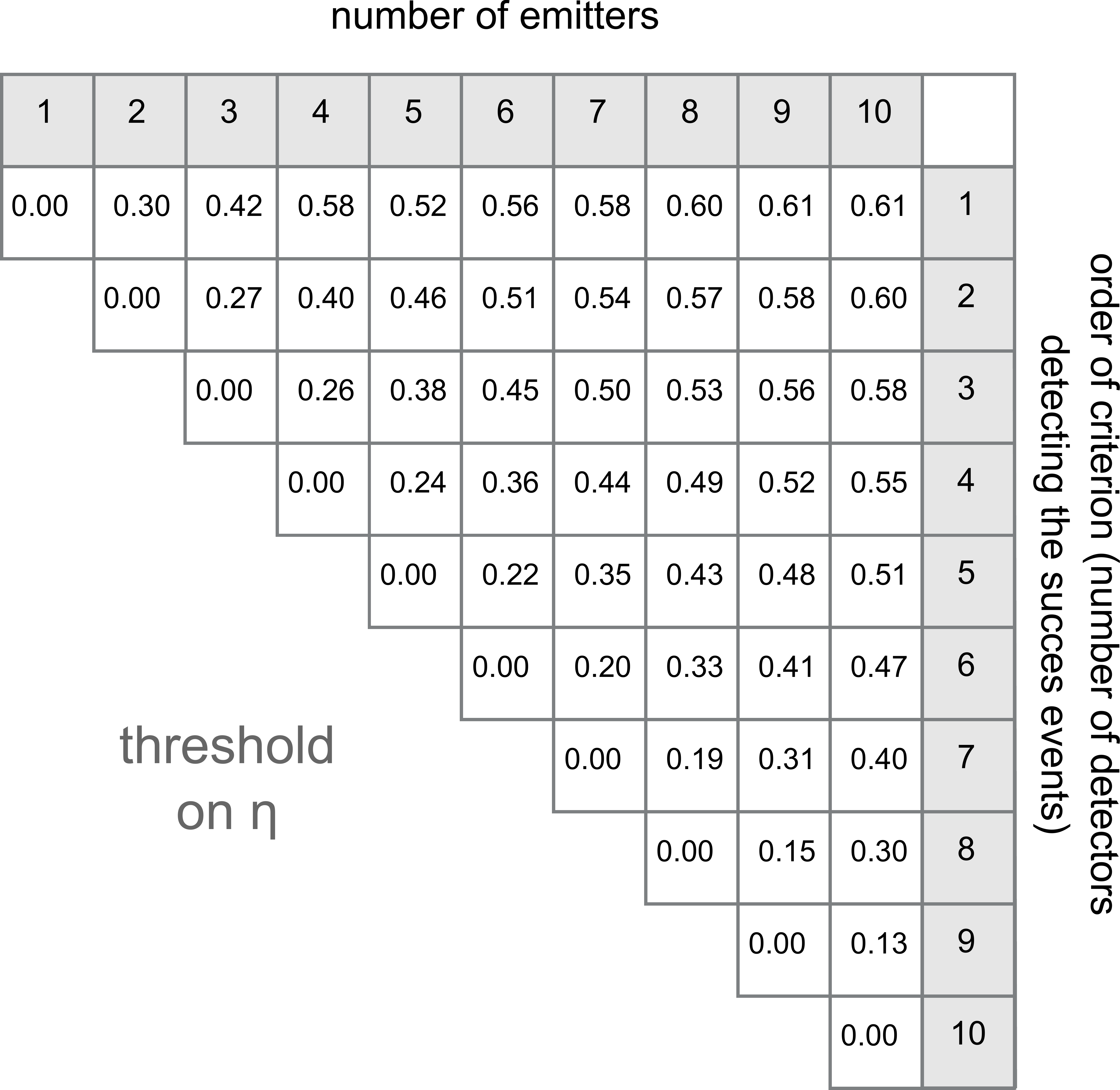}}
\caption{Table of minimal efficiencies of detection $\eta$ that render to detect quantum non-Gaussianity from an ensemble of ideal single photon emitters utilizing a specific order of a criterion. It demonstrates the quantum non-Gaussianity of the emitted light is visible for arbitrarily efficiency $\eta>0$ only if the order agrees with number of emitters presented in the source.}
\label{eta}
\end{figure}

A classical background noise contaminates the signal from an ensemble of emitters by false clicks. The photon statistics of the multi-mode noise is Poissonian in common experimental platforms, we therefore focus on this case. If emission of each emitter is spoiled by noise always with a mean photon number $\bar{n}$, the mean photon number of the total noise increases with number of emitters $\langle n \rangle = m \bar{n}$ and affects the no-click probabilities so this
\begin{equation}
R_{0,n}=\left(1-\frac{\eta n}{N}\right)^m \exp(-m\bar{n}n/N).
\label{R0SN}
\end{equation}
We identify the criterion order with number of emitters in the following discussions. The noise has a strong impact on the observation of quantum non-Gaussianity as depicted in Fig.~4. In the limit of very attenuated states, the probabilities of success and error get approximations
\begin{eqnarray}
R_m &\approx &\frac{m!}{(m+1)^m}\eta^m \nonumber \\
R_{m+1} &\approx &\frac{m!}{(m+1)^{m}}\eta^m m\bar{n},
\label{RnApp}
\end{eqnarray}
where $m$ denotes the number of single photon emitters and $m \bar{n} \ll \eta$. Thus, the approximate criteria in Eq. (\ref{approx}) require in this regime
\begin{equation}
\eta>\frac{H_m^{2/m}(x)}{\sqrt[m]{m!}}\sqrt{\frac{m \bar{n}}{2(m+1)}},
\label{appS}
\end{equation}
where $x$ satisfies $H_{m+1}(x)=0$ and $m\bar{n}\ll \eta$. Figure~4 presents the convergence of this condition to numerically derived thresholds on $\eta$. The sensitivity of quantum non-Gaussianity to noise increases with greater number of emitters, therefore any experimental tests will require low noise for large $m$. 
 \begin{figure}[t]
\centerline{\includegraphics[width=9cm]{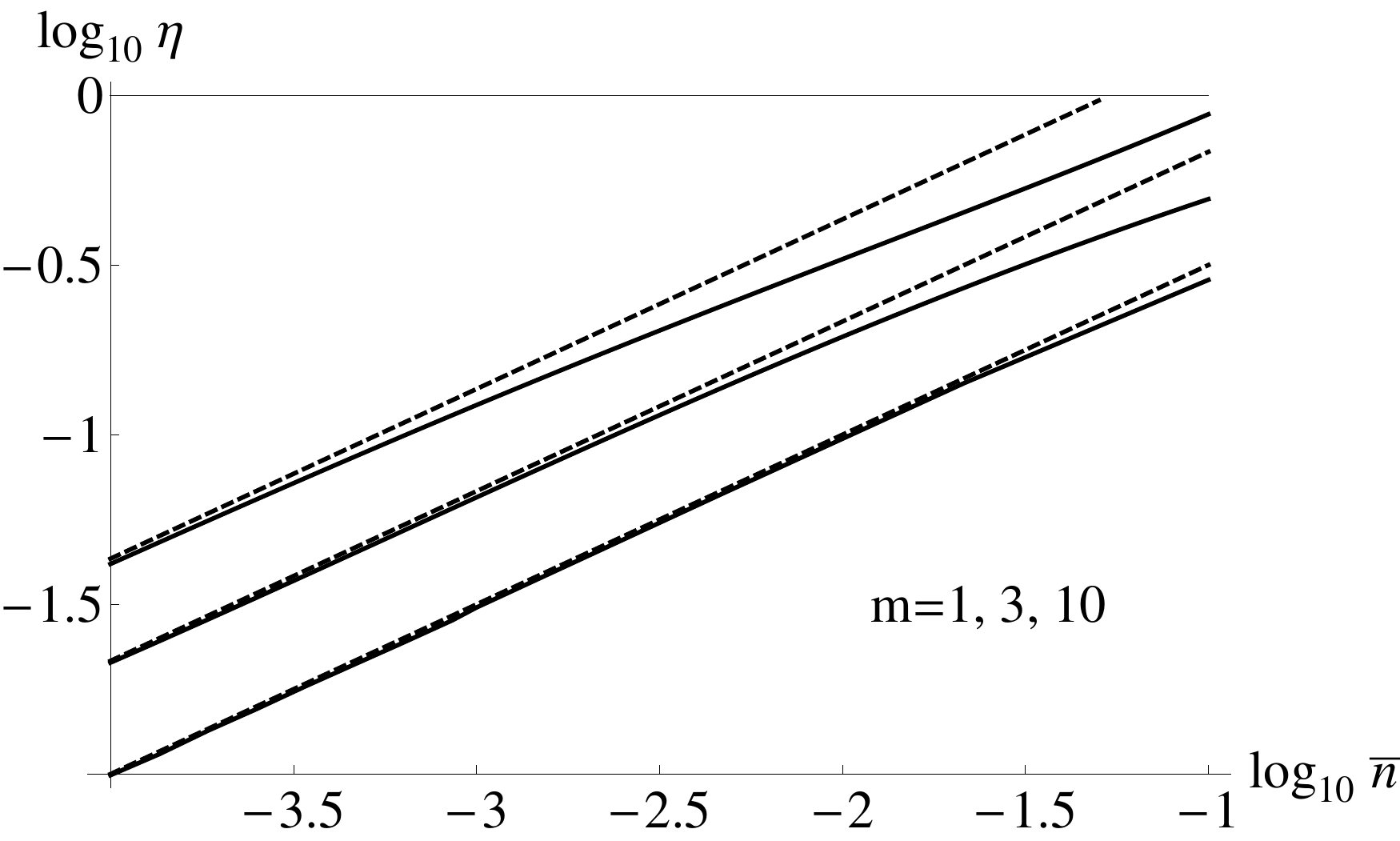}}
\caption{The influence of background noise with Poissonian statistics on the visibility of quantum non-Gaussianity is drawn by solid lines for number of emitters $m=1,\ 3$ and $10$. The higher lines corresponds to the greater number of emitters. The dashed lines appropriate to the threshold efficiency derived from approximate condition in Eq. (\ref{approx}).}
\label{source}
\end{figure}
 \begin{figure}[t]
\centerline{\includegraphics[width=9cm]{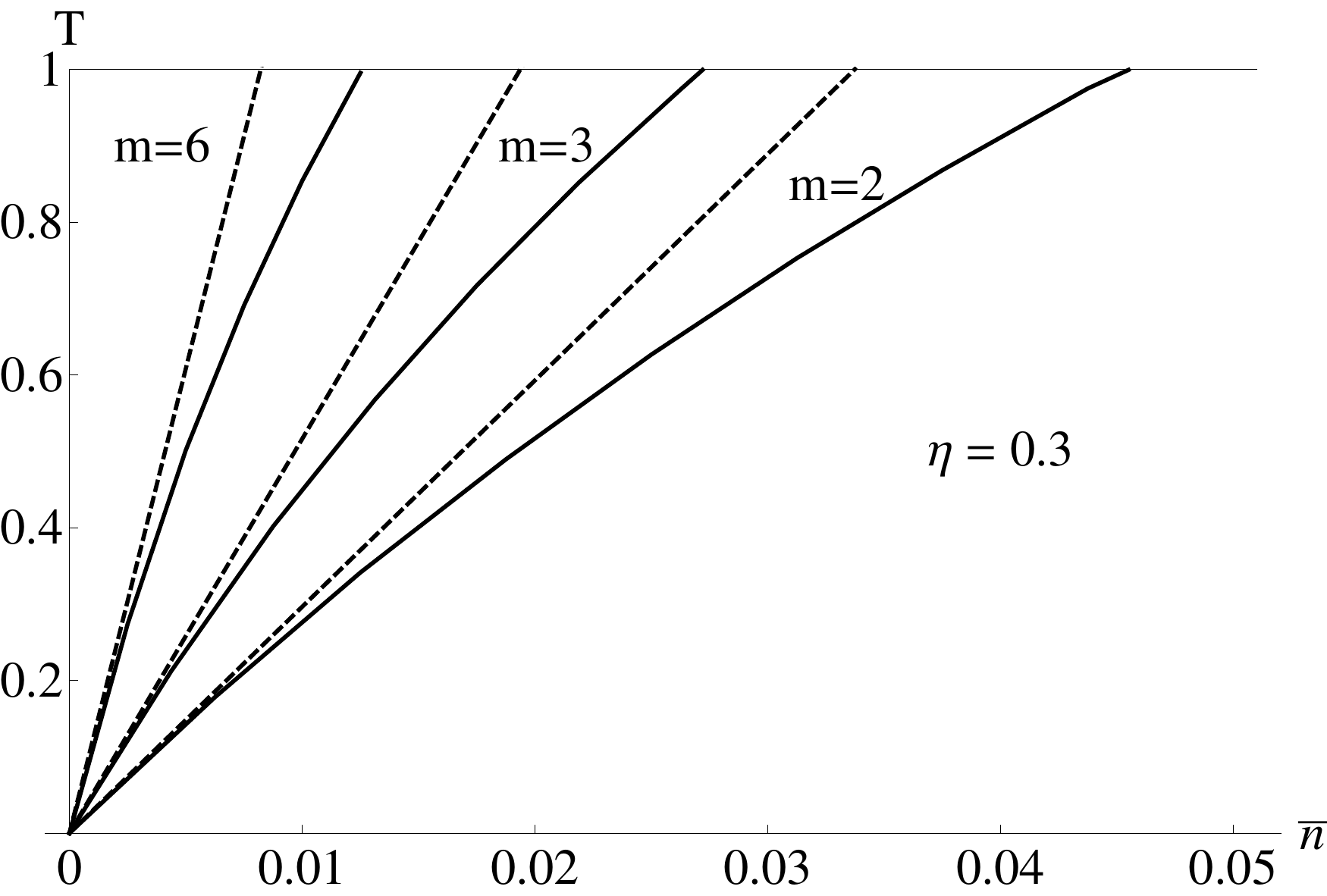}}
\caption{The robustness of quantum non-Gaussianity against losses is depicted by solid lines. The simulated state has efficiency $\eta=0.3$ and consists of $m=2,\ 3$ and $6$ emitters. The dashed lines appropriate to a limitation on efficiency $T$ yielded from an approximation $\bar{n} T \ll 1$.  Coincidence of full and dashed lines confirms validity of our approximation for strongly attenuated very good single photon emitters. }
\label{at}
\end{figure}

The quantum non-Gaussianity is influenced by losses and by quantum efficiency of individual APDs. If the detector has equal quantum efficiency in each channel or if the variation is compensated with appropriate splitting ratios, the quantum efficiency can be involved in the lossy channel. Otherwise the task has to be treated such as in Ref. \cite{me3}.
Quantum non-Gaussianity of attenuated single photon states is very loss-tolerant \cite{Ivo}.
In the case of multi-photon light, an attenuating channel with efficiency of transmission $T$ decreases the efficiency of detection $\eta \rightarrow T \eta $ of the signal and mean photon number presented in the noise $\bar{n} \rightarrow \bar{n}T$. Figure~5 depicts the robustness attributed to a state, in which signal with $\eta=0.3$ is aggravated by noise. If the state has subsided mean number of photon sufficiently, the quantum non-Gaussianity tolerate loss-channels with
\begin{equation}
T> \frac{m \bar{n} H_{m}^{4/m}(x) }{2\eta^2 (m+1)(m!)^{2/m} },
\end{equation}
where it is supposed $\bar{n} T \ll 1$ and $x$ satisfies $H_{m+1}(x)=0$.
As number $m$ of emitters increases, loss-tolerance requires much smaller $\bar{n}$ of noise. It is therefore important to reduce background noise even more for experimental verification of quantum non-Gaussianity from many emitters.
 \begin{figure}[t]
\centerline{\includegraphics[width=9cm]{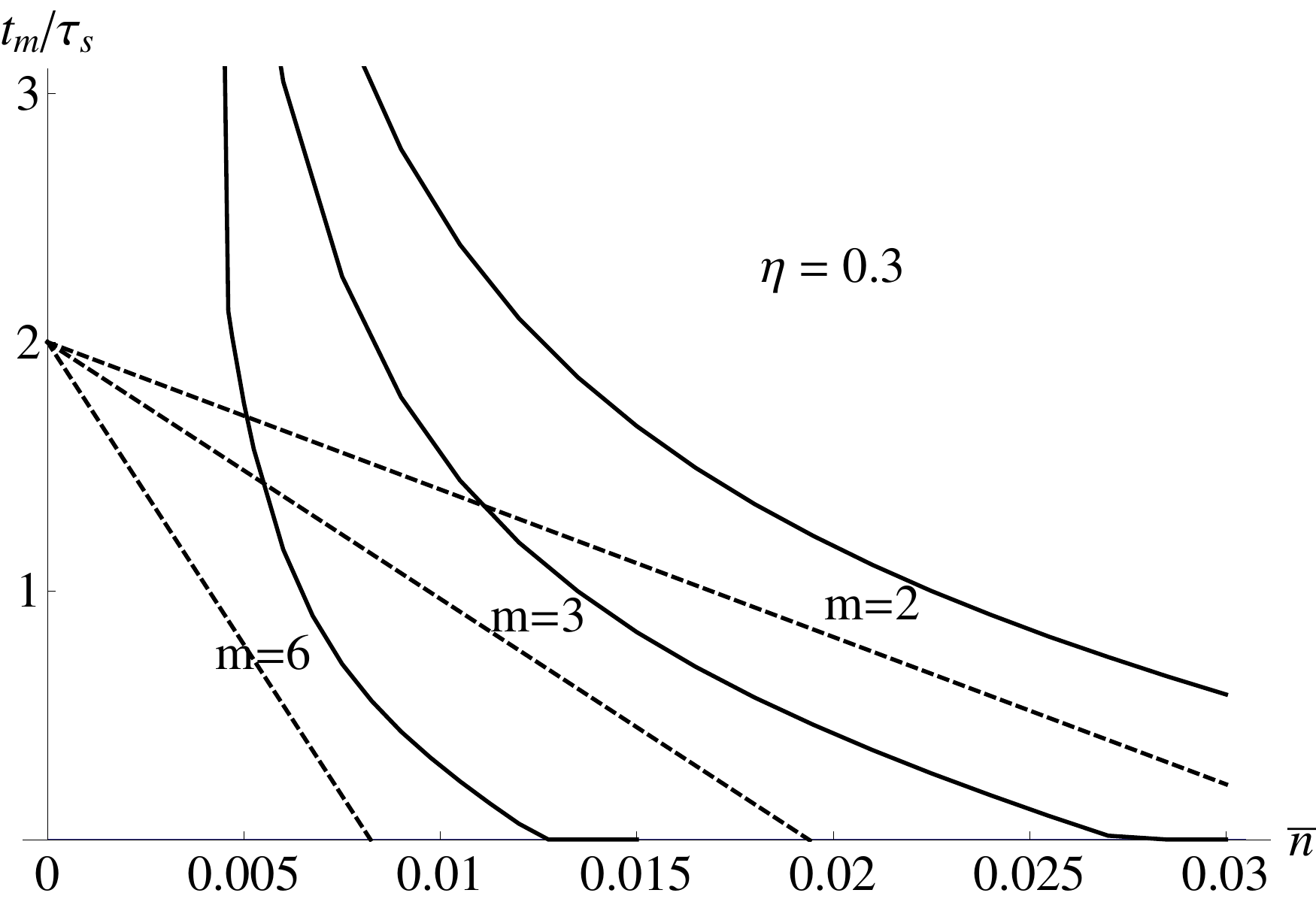}}
\caption{The influence of escaping of emitters from the source on quantum non-Gaussianity. The consider source consists of $m=2,\ 3$ and $6$ independent emitters with efficiency of emission $\eta=0.3$. Each emitter has a mean storage time $\tau_s$ in a source and is contaminated by a noise with mean number of photons $\bar{n}$. The quantum non-Gaussianity is observable in measurement with duration $t_m$ shorter then time given by the solid lines. The dashed lines appropriate to estimation of safe measurement duration based on the approximate formulas.}
\label{gauss}
\end{figure}

Another imperfection of a source consisting of many emitters is escaping of the emitters to an environment. Consider, an emitter can be observed with a probability decaying exponentially in time $P(t)=\exp(-t/\tau_s)$, where $\tau_s$ is storage time of the emitter. In this case, the measurement has to be repeated several times with the same initial number of emitters to get a trustworthy statistics of clicks. Thus, the no-clicks probabilities gain
\begin{equation}
\langle R_{0,k} \rangle= \int_{t=t_0}^{t_0+t_M} \left(1-\frac{\eta k}{N}e^{-t/\tau_s}\right)^m e^{- \frac{m k}{N} \bar{n}\exp(-t/\tau_s)} \mathrm{d}t,
\label{R0Aver}
\end{equation}
where $\langle ... \rangle$ means average over several measurements with the same initial number of emitters and $t_M$ is measurement duration. If the mean number of photons of the noise is negligible, the quantum non-Gaussianity is still visible for an arbitrary ensemble of single photon emitters with $\eta>0$, that implies from the absence of error events. However, if the emitters escape significantly the hierarchy puts a stricter condition on mean photon number of the noise as depicted in Fig.~6. The estimation of the maximal time enabling the detection in the limit of strongly attenuated state $\bar{n} \ll 1$ is given by
\begin{equation}
\frac{2t_M}{\tau_s}\eta^2 <\eta^2-\frac{m \bar{n} H_{m}^{4/m}(x) }{4 (m+1)(m!)^{2/m} },
\end{equation}
where $H_{m+1}(x)=0$ and it is supposed $\frac{1}{12}\left(\frac{t_M}{\tau_s}\right)^2 \ll 1$ and $\frac{m^3}{1440} \left(\frac{t_M}{\tau_s}\right)^4 \ll 1$. It issues from averaging approximation of success and error probabilities (\ref{RnApp}) and expansion of the derived condition for small times. It says how much the left side of the condition in Eq. (\ref{appS}) has to be  greater than the right side so that the quantum non-Gaussianity is observable, when this imperfection of the source is significant.

\section{Conclusion and experimental outlook}
The detection of quantum non-Gaussianity of light radiated from a large ensemble of single photon emitters requires adequately numerous array of detectors or calibrated detection using intensified CCD camera. The main limitation for this detection is given by noise,
which has to be suppressed more for higher number of emitters. On the other hand, quantum non-Gaussianity of multi-photon light tolerates decaying number of emitters in the source and losses above fifty percent and therefore is suitable for realistic experimental platforms. The criteria of quantum non-Gaussianity represent a useful tool for analysis of a cluster of single photon emitters, production of Fock states of light and can be extended to another systems like phonons in optomechanics. The ideal platform for the test are ions in a crystal \cite{me}, where quality of the single photon emitters, their stability, low background noise and sufficient collection efficiency enable to test quantum non-Gaussianity for high number of emitters. Such advanced test will extend recent experimental demonstration of quantum non-Gaussian light from single trapped ion \cite{Daniel}. 

\section*{Funding}
Czech Science Foundation GB14-36681G;
financial support of Palacký University IGA-Prf-2016-009.



\end{document}